\documentclass[runningheads]{llncs}
\usepackage{graphicx}
\usepackage{subfigure}
\usepackage{algorithm}
\usepackage{algpseudocode}
\usepackage{multirow}
\begin{document}
\title{PlutoNet: An Efficient Polyp Segmentation Network with Modified
Partial Decoder and Decoder Consistency Training}

\author{Tugberk Erol\inst{1}\orcidID{0000-0002-0831-7307} \and
Duygu Sarikaya\inst{2}\orcidID{0000-0002-2083-4999}}
\authorrunning{Erol et al.}
%
\institute{Computer Engineering Department,  Graduate School of Natural and Applied Sciences, Gazi University, Ankara, 06570, Turkey  \and School of Computing, University of Leeds, Leeds, LS2 9JT, UK
}

%
%

%
\maketitle              
%

%
%

\begin{abstract}
 Deep learning models are used to minimize the number of polyps that goes unnoticed by the experts and to accurately segment the detected polyps during interventions. Although state-of-the-art models are proposed, it remains a challenge to define representations that are able to generalize well and that mediate between capturing low-level features and higher-level semantic details without being redundant. Another challenge with these models is that they require too many parameters, which can pose a problem with real-time applications. To address these problems, we propose PlutoNet for polyp segmentation which requires only 2,626,537 parameters, less than 10\% of the parameters required by its counterparts. With PlutoNet, we propose a novel \emph{decoder consistency training} approach that consists of a shared encoder, the modified partial decoder which is a combination of the partial decoder and full-scale connections that capture salient features at different scales without being redundant, and the auxiliary decoder which focuses on higher-level relevant semantic features. We train the modified partial decoder and the auxiliary decoder with a combined loss to enforce consistency, which helps improve the encoder’s representations. This way we are able to reduce uncertainty and false positive rates. We perform ablation studies and extensive experiments which show that PlutoNet performs significantly better than the state-of-the-art models, particularly on unseen datasets and datasets across different domains.
 
\keywords{Polyp Segmentation  \and Consistency Training \and Modified Partial Decoder}
\end{abstract}

\section{Introduction} \label{Introduction}

Studies show that during colonoscopy, depending on their type and size, 14-30\% of polyps go unnoticed by the experts \cite{jha2020sessile}. Deep learning models are used to minimize the number of polyps that goes unnoticed by the experts and to accurately segment the detected polyps during these interventions. It remains a challenge, however, for these models to generalize to different domains and to learn representations that mediate between capturing low-level features and higher-level semantic details without being redundant. 
\par
\begin{figure}[]
 
    \includegraphics[width=1.0\textwidth]{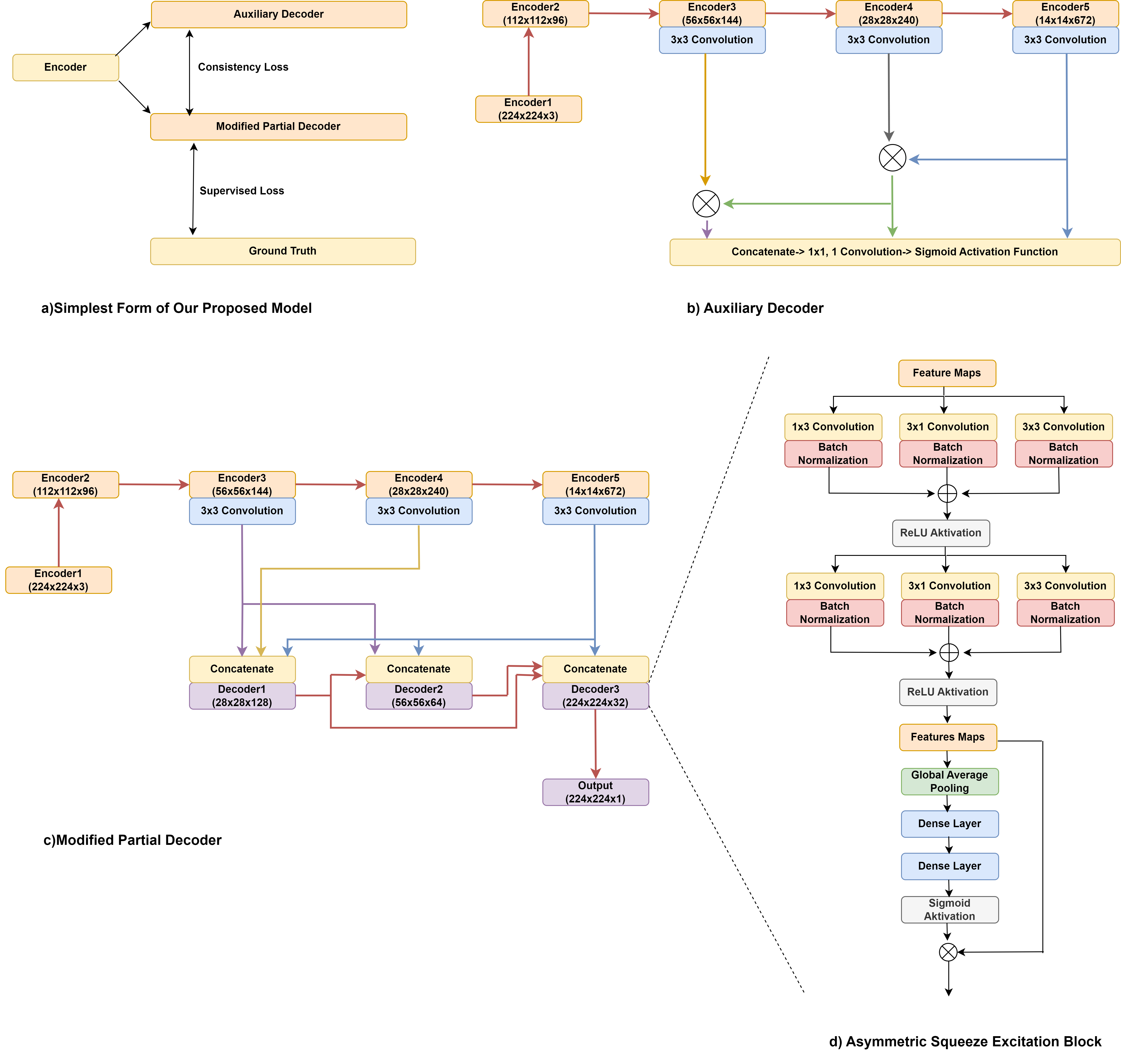}
    \caption{The overview of PlutoNet. In the top left corner (a), a simple representation of our proposed model is shown. In PlutoNet, a shared encoder, the modified partial decoder, and the auxiliary decoder are trained with a combined loss to enforce consistency. In the top right corner (b), the auxiliary decoder is shown. It carries out elementwise multiplication of higher-level encoders and then concatenates them. In the bottom left corner (c), the modified partial decoder, which is a combination of the partial decoder and full-scale connections is shown. In the bottom right corner (d), the details of the decoder layers which use a combination of asymmetric convolutions and a squeeze and excitation block are shown. }
    \label{overviewPlutoNet}
\end{figure}

Most state-of-the-art models rely on an encoder-decoder structure with symmetrical contracting (down-sampling) and expansive (up-sampling) paths, and skip connections, to capture both low and high-level information \cite{unet}. As repeated feature down-sampling may cause small polyps to be easily degraded \cite{shallowattention}, we wish to carry as much low and high-level information as possible through skip connections. However, this leads these state-of-the-art models to carry redundant information and increases the number of parameters required, which can pose a problem with real-time applications. Moreover, the low-level features may cause noise, making the segmentation task challenging \cite{shallowattention}. Therefore, we need to learn representations that mediate between capturing low-level features and higher-level semantic details without being redundant. To address this need, we propose PlutoNet which requires only 2,626,537 parameters. With PlutoNet, we propose a novel \emph{decoder consistency training} approach, which ensures a balance between the salient details at different scales learned through the modified partial decoder and the more relevant higher-level semantic features learned through the auxiliary decoder. PlutoNet adopts a lightweight encoder-decoder structure~\cite{unet} and extends on the \emph{modified partial decoder} \cite{efficient}, which is a combination of partial decoder \cite{pd} and full-scale connections \cite{unet3plus}. Using \emph{modified partial decoder}, we are able to reduce the number of parameters by ignoring skip connections to the low-level features which may be redundant. Polyps in colonoscopy images have varying sizes, appearances, and aspect ratios. In order to handle these variations, we use asymmetric convolutions, and then we increase the representation of the more relevant features by weighting each feature map using a squeeze and excitation block as proposed by Erol et al.\cite{efficient}. Then we enforce consistency by combining the loss of the modified partial decoder and the auxiliary decoder, which encourages the predictions of the decoders to be consistent, filters out irrelevant salient features, and helps improve the encoder’s representations. This way we are able to better focus on the polyps, and reduce uncertainty and false positive rates. The auxiliary decoder adds only $200$ parameters to our network architecture and is only needed for training. An overview of our model is demonstrated in Figure \ref{overviewPlutoNet}.

%
%

%
%

\par
The main contributions of this paper are: (1) the novel \emph{decoder consistency training} approach that ensures a balance between the salient details at different scales learned through the modified partial decoder and the more relevant higher-level semantic features learned through the auxiliary decoder. Although conventionally used in unsupervised learning problems, we show that the representations learned through consistency training combining the loss of differently specialized decoders perform well in segmentation tasks. To our knowledge, our study is the first to propose decoder-level consistency training between two decoders with different focuses on learning more salient features and higher-level semantic features. (2) We present PlutoNet which requires only 2,626,537 parameters, which is far fewer than the state-of-the-art models, about less than 10\% of the parameters required by its counterparts. In order to achieve this, we adopt a lightweight encoder-decoder structure~\cite{unet} and extend on the \emph{modified partial decoder} \cite{efficient} that reduces the number of parameters by ignoring skip connections to the low-level features which may be redundant. The auxiliary decoder adds only $200$ parameters to our network architecture and is only needed for training. (3) We tested our model extensively for the segmentation of polyps in colonoscopy and wireless endoscopy images on five different public datasets. PlutoNet performs significantly better than the state-of-the-art models, particularly on unseen datasets and datasets across different domains, which demonstrates its generalizability. (4) We carried out ablation studies to show the effectiveness of our consistency training approach.

%
%

\section{Related Work} \label{RelatedWork}

Huang et al.~\cite{unet3plus} proposed UNET 3+, a U-Net based architecture with full-scale connections which integrate low and high-level information at different scales to minimize the loss of information. Wu et al. ~\cite{pd}'s experiments showed that the third encoder layer carried low-level features as well as high-level ones, therefore concatenations of lower layers are mostly redundant. They also showed that although lower layers require more computation, they contribute less to the overall performance. Based on these findings, they developed the partial decoder. Wei et al. ~\cite{shallowattention} proposed a novel polyp segmentation network titled Shallow Attention Network following the findings of Wu et al. ~\cite{pd}. Erol et al. ~\cite{efficient} proposed to use a combination of the partial decoder and full-scale connections.

Consistency training has been used in semi-supervised learning to leverage unlabeled data by creating variations of the available data and combining the loss with the loss that comes from training the data available. Ouali et al. ~\cite{Ouali} proposed cross-consistency training which improves the encoder's representations through different perturbations for semi-supervised semantic segmentation. Sohn et al. \cite{fixmatch} presented consistency training for image classification. They used pseudo labels with weak and strong augmentations of images. More recently, Wu et al. \cite{mutual} proposed mutual consistency learning for semi-supervised medical image segmentation. They used a shared encoder and almost identical decoders with three different up-sampling strategies. In this work, we propose a novel \emph{decoder consistency training} approach, which ensures a balance between the salient details learned through the modified partial decoder and the higher-level semantic features learned through the auxiliary decoder.

%
%

\section{Method}
An overview of our model is demonstrated in Figure \ref{overviewPlutoNet}. We primarily adopt a lightweight encoder-decoder structure using the last three encoder layers of EfficientNetB0, followed by the \emph{modified partial decoder} \cite{efficient}. We apply $64$ convolution filters to the output of the encoder layers before they go into the full-scale connections, which further reduces the number of parameters. In order to handle variations in appearance, we use asymmetric convolutions. Each decoder layer consists of an asymmetric convolution block followed by a squeeze and excitation block. Then we enforce consistency by combining the loss of the modified partial decoder and the auxiliary decoder.

%
%

\subsection{Modified Partial Decoder}

Based on the findings of the experiments by Wu et al. \cite{pd}, Erol et al. \cite{efficient} removed the full-scale skip connections of the earlier layers, $e^1$ and $e^2$. This way, they combined partial decoder and full-scale skip connections, namely the \textit{modified partial decoder} at different scales, while reducing the redundant and less informative features of the earlier layers, $e^1$ and $e^2$.

\begin{equation}
\label{denklemacb}
    acb \gets relu(bn(conv(3x1)) + bn(conv(1x3)) + bn(conv(3x3)))
\end{equation}

 \begin{equation}
 \centering
\label{denklemd1}
d1 \gets se(acb(c(conv(e^{3}), conv(e^{4}), conv(e^{5}))))
\end{equation}

 \begin{equation}
 \centering
\label{denklemd2}
d2 \gets se(acb(c(d^{1}, conv(e^{3}), conv(e^{5}))))
\end{equation}

 \begin{equation}
 \centering
\label{denklemd3}
d3 \gets se(acb(c(d^{1}, d^{2}, conv(e^{5}))))
\end{equation}

In Equation \ref{denklemd1}, \ref{denklemd2} and \ref{denklemd3}, c, $d^1$, $d^2$, $d^3$, $e^3$, $e^4$, $e^5$, $se$, $acb$ represent  concatenate, decoder1, decoder2, decoder3, encoder3, encoder4, encoder5, squeeze and excitation and asymmetric convolution block, respectively. As mentioned earlier, we skip the connections to $e^1$ and $e^2$ as the higher layers carry the low-level features that are learned through the earlier layers which makes the connections to the two early layers redundant. $e^3$ and $e^4$ concatenate with the same and larger scale feature maps. $e^5$ is concatenated with all of the three decoder layers. We also concatenate inter-decoder layers at smaller and larger scales. These connections are demonstrated in Figure ~\ref{overviewPlutoNet}. Ding et al~\cite{asymmetric} proposed asymmetric convolutions to strengthen kernels, making them able to handle variations in appearance and size. In our work, we use asymmetric convolutions to handle variations in appearance, aspect ratio, and size of the polyps as suggested by Erol et al. \cite{efficient}. After we enrich the feature space using asymmetric convolutions, we weigh each feature map using a squeeze and excitation block to increase the representation of the more relevant features. This channel-wise feature recalibration is done at every layer. A detailed view of the Asymmetric Convolution block and the Squeeze and Excitation Block can be seen in Figure ~\ref{overviewPlutoNet}. Equation \ref{denklemacb} shows the asymmetric convolution block structure. $bn$ and $conv$ represent batch normalization and convolution.

\subsection{Decoder Consistency Training}

We propose a novel consistency training approach that consists of a shared encoder, the modified partial decoder, and the auxiliary decoder that are trained with a combined loss to enforce consistency (Figure\ref{overviewPlutoNet}). While conventionally used in unsupervised segmentation problems \cite{Ouali}, we use our consistency training approach to ensure a balance between the salient details at different scales learned through the modified partial decoder and the more relevant higher-level semantic features learned through the auxiliary decoder. For the auxiliary decoder, we use the Shallow Attention proposed by Wei et al.\cite{shallowattention}. While the modified partial decoder learns salient details at different scales, from an enriched feature space extracted using asymmetric convolutions, the auxiliary decoder focuses on more relevant higher-level semantic features learned through an attention mechanism built on a series of element-wise multiplications of features extracted only from higher layers. Equation \ref{shallow_equation} shows how the auxiliary decoder works.

 \begin{equation}
\centering
\label{shallow_equation}
               d \gets conv(c(e^{3}*e^{4}*e^{5}, e^{4}*e^{5}, e^{5}))
\end{equation}

\begin{algorithm}

\caption{Consistency Training Algorithm}

 $P_m \gets f_m(x, \theta)$ \Comment{$f_m$ and $P_m$ represent main decoder and its prediction.}
 
 $P_a \gets f_a(x, \theta)$ \Comment{$f_a$ and $P_a$ represent auxiliary decoder and its prediction.}
 
 $L_c = 2*(1-\frac{\sum P_{m} * P_{a}}{\sum P^2_{m} + \sum P^2_{a} + \epsilon})$ \Comment{Consistency Loss}
 
 $L_s = 2*(1-\frac{\sum P_{m} * P_{t}}{\sum P^2_{m} + \sum P^2_{t} + \epsilon})$ \Comment{Supervised Loss}
 
 $L = L_s + \alpha L_c$ \Comment{Total Loss}
\label{algoritma}

\end{algorithm}

We enforce consistency by combining the loss of the modified partial decoder and the auxiliary decoder, which encourages the predictions of the decoders to be consistent. The algorithm we follow is shown in Algorithm \ref{algoritma}, it also shows how we calculate the total loss, where $P_{t}$, $P_{m}$ and $P_{a}$ represents ground truth, the output of the modified partial decoder and the output of the auxiliary decoder, respectively. The auxiliary decoder adds only $200$ parameters to our network architecture and is only needed for training.

\section{Experimental Details} \label{experimental-details}

We evaluated our model extensively for the segmentation of polyps in colonoscopy and wireless endoscopy images on five different public datasets and carried out an ablation study to show the effectiveness of our decoder consistency training approach. We followed the experimentation set-up suggested by Fan et al. ~\cite{pranet}; we split Kvasir-SEG ~\cite{jha2020sessile} and CVC-ClinicDB~\cite{PMID:25863519} datasets as 80\% training, 10\% validation, and 10\% testing, and carried out ablation studies on the Kvasir-SEG dataset. Then we tested our model further on ETIS~\cite{etis}, Endoscene~\cite{endoscene}, and CVC-ColonDB~\cite{colondb} datasets. We implemented our model in TensorFlow accelerated by NVIDIA RTX 3050TI 4GB. All images are resized to $224x224x3$. We used random rotation and horizontal flip data augmentation techniques. We set up an early stopping scheme according to the validation loss (trained for $30$ epochs). We set the initial learning rate to $1e-4$ and used the Adam optimizer. 

\section{Results} \label{results}

We compared PlutoNet's performance to a benchmark consisting of the state-of-the-art models, namely, UNet~\cite{unet}, UNet++ ~\cite{unet++}, SFA~\cite{sfanet}, PraNet ~\cite{pranet}, MSNet~\cite{MSNet} and Shallow Attention~\cite{shallowattention}. Table \ref{5_dataset_data} shows our model's results compared to the results of the benchmark studies. A comparison of the number of parameters each benchmark model requires is also shown. 

\begin{table*}[htbp]

\caption{A comparison of our model's performance to the state-of-the-art polyp segmentation models is demonstrated.}

\begin{tabular*}{\textwidth}{l|ll|ll|ll|ll|ll|c}

\hline
\multirow{2}{*}{Methods}  & \multicolumn{2}{l|}{Kvasir}                          & 

\multicolumn{2}{l|}{ClinicDB}                        & \multicolumn{2}{l|}{ColonDB}                         & \multicolumn{2}{l|}{EndoScene}                       & \multicolumn{2}{l|}{ETIS}                            & \multicolumn{1}{l}{Param} \\ \cline{2-12} 
                           & \multicolumn{1}{l|}{Dice}           & IoU            & \multicolumn{1}{l|}{Dice}           & IoU            & \multicolumn{1}{l|}{Dice}           & IoU            & \multicolumn{1}{l|}{Dice}           & IoU            & \multicolumn{1}{l|}{Dice}           & IoU            &                        \\ \hline
UNet              & \multicolumn{1}{l|}{0.818}          & 0.746          & \multicolumn{1}{l|}{0.823}          & 0.755          & \multicolumn{1}{l|}{0.512}          & 0.444          & \multicolumn{1}{l|}{0.710}          & 0.627          & \multicolumn{1}{l|}{0.398}          & 0.335          & 15.7M                                     \\ \hline
UNet++            & \multicolumn{1}{l|}{0.821}          & 0.743          & \multicolumn{1}{l|}{0.794}          & 0.729          & \multicolumn{1}{l|}{0.483}          & 0.410          & \multicolumn{1}{l|}{0.707}          & 0.624          & \multicolumn{1}{l|}{0.401}          & 0.344          & 9M                                          \\ \hline
SFA             & \multicolumn{1}{l|}{0.723}          & 0.611          & \multicolumn{1}{l|}{0.700}          & 0.607          & \multicolumn{1}{l|}{0.469}          & 0.347          & \multicolumn{1}{l|}{0.467}          & 0.329          & \multicolumn{1}{l|}{0.297}          & 0.217          & -                                          \\ \hline
PraNet            & \multicolumn{1}{l|}{0.898}          & 0.840          & \multicolumn{1}{l|}{0.899}          & 0.849          & \multicolumn{1}{l|}{0.709}          & 0.640         & \multicolumn{1}{l|}{0.871}          & 0.797          & \multicolumn{1}{l|}{0.628}          & 0.567          & 30.5M                                       \\ \hline
MSNET              & \multicolumn{1}{l|}{\textbf{0.907}} & \textbf{0.862} & \multicolumn{1}{l|} {\textbf{0.921}}          & \textbf{0.879} & \multicolumn{1}{l|}{\textbf{0.755}} & \textbf{0.678} & \multicolumn{1}{l|}{0.869}          & 0.807          & \multicolumn{1}{l|}{0.719}          & 0.664          & 25.2M                                      \\ \hline
SANet & \multicolumn{1}{l|}{0.904}          & 0.847          & \multicolumn{1}{l|}{0.916}          & 0.859          & \multicolumn{1}{l|}{0.753}          & 0.670          & \multicolumn{1}{l|}{0.888}          & 0.815 & \multicolumn{1}{l|}{0.750} & 0.654 & 23.9M                                      \\ \hline
Ours                       & \multicolumn{1}{l|}{0.895}          & 0.811          & \multicolumn{1}{l|}{0.909} & 0.832          & \multicolumn{1}{l|}{0.694}          & 0.531          & \multicolumn{1}{l|}{\textbf{0.919}} & {\textbf{0.851}}          & \multicolumn{1}{l|}{\textbf{0.829}}          & {\textbf{0.709}}          & \textbf{2.6M}                               \\ \hline
\end{tabular*}

\label{5_dataset_data}
\end{table*}

Our model outperformed UNet, UNet++, and SFA on all datasets for Dice and IoU metrics. Even though we only used about less than 10\% of the parameters required by PraNet, MSNet, and Shallow Attention, our model outperformed state-of-the-art models on ETIS with an 82.9\% Dice score and on Endoscene with a 91.9\% Dice score. It should be noted that PlutoNet performs significantly better than the state-of-the-art models on unseen datasets. Among these unseen datasets; ETIS~\cite{etis}, Endoscene~\cite{endoscene}, and CVC-ColonDB~\cite{colondb}, PlutoNet outperforms all benchmark models on two of them for both Dice and IoU metrics. Particularly, PlutoNet outperforms all other models with a large margin on ETIS, which is a dataset of images captured by capsule endoscopy and differs greatly in resolution. This supports PlutoNet's ability to learn stronger representations that generalize well to unseen datasets and datasets across different domains.

\subsection{Ablation Study} \label{ablation_study}

\begin{figure}[ht]

    \subfigure{\includegraphics[width=0.09\textwidth]{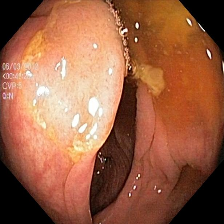}}
    \hspace{-0.18cm}
    \subfigure{\includegraphics[width=0.09\textwidth]{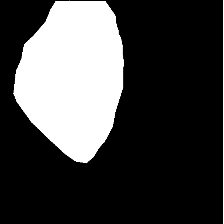}}
    \hspace{-0.18cm}
    \subfigure{\includegraphics[width=0.09\textwidth]{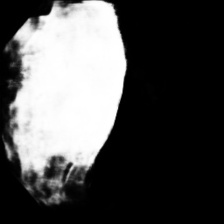}}
    \hspace{-0.18cm}
    \subfigure{\includegraphics[width=0.09\textwidth]{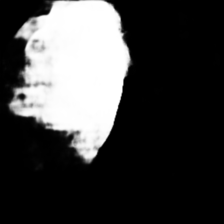}}
    \hspace{-0.18cm}
    \subfigure{\includegraphics[width=0.09\textwidth]{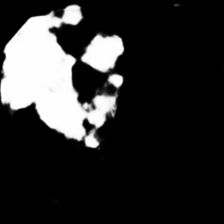}}
    \hspace{-0.18cm}
    \subfigure{\includegraphics[width=0.09\textwidth]{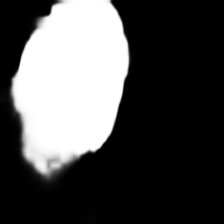}}
    \hspace{-0.18cm}
    \subfigure{\includegraphics[width=0.09\textwidth]{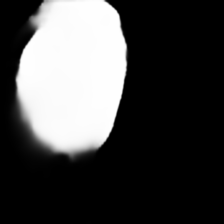}}
    \hspace{-0.18cm}
    \subfigure{\includegraphics[width=0.09\textwidth]{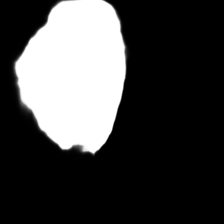}}
    \hspace{-0.18cm}
    \subfigure{\includegraphics[width=0.09\textwidth]{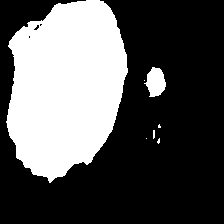}}
    \hspace{-0.18cm}
    \subfigure{\includegraphics[width=0.09\textwidth]{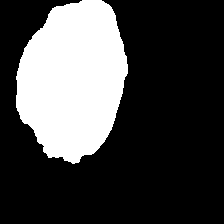}}
    \vspace{-0.35cm}

    \subfigure{\includegraphics[width=0.09\textwidth]{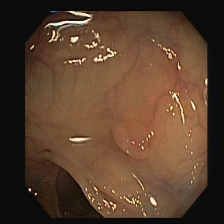}}
    \hspace{-0.18cm}
    \subfigure{\includegraphics[width=0.09\textwidth]{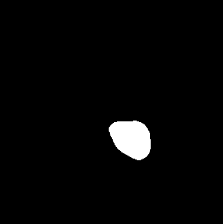}}
    \hspace{-0.18cm}
    \subfigure{\includegraphics[width=0.09\textwidth]{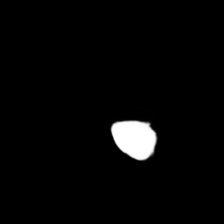}}
    \hspace{-0.18cm}
    \subfigure{\includegraphics[width=0.09\textwidth]{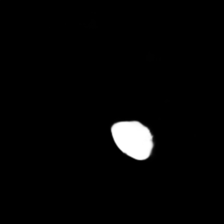}}
    \hspace{-0.18cm}
    \subfigure{\includegraphics[width=0.09\textwidth]{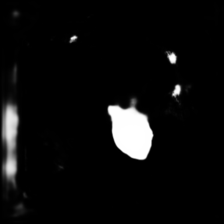}}
    \hspace{-0.18cm}
    \subfigure{\includegraphics[width=0.09\textwidth]{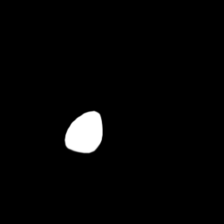}}
    \hspace{-0.18cm}
    \subfigure{\includegraphics[width=0.09\textwidth]{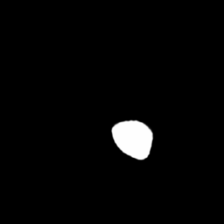}}
    \hspace{-0.18cm}
    \subfigure{\includegraphics[width=0.09\textwidth]{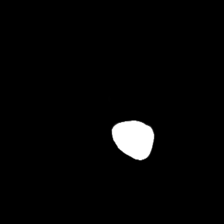}}
    \hspace{-0.18cm}
    \subfigure{\includegraphics[width=0.09\textwidth]{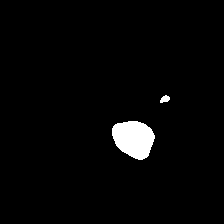}}
    \hspace{-0.18cm}
    \subfigure{\includegraphics[width=0.09\textwidth]{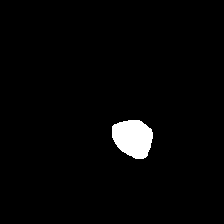}}
    \hspace{-0.18cm}
     \vspace{-0.35cm}

    \subfigure{\includegraphics[width=0.09\textwidth]{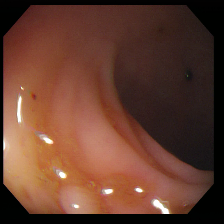}}
    \hspace{-0.18cm}
    \subfigure{\includegraphics[width=0.09\textwidth]{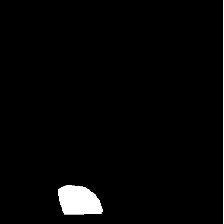}}
    \hspace{-0.18cm}
    \subfigure{\includegraphics[width=0.09\textwidth]{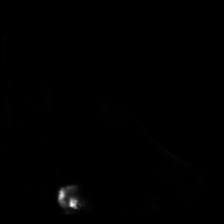}}
    \hspace{-0.18cm}
    \subfigure{\includegraphics[width=0.09\textwidth]{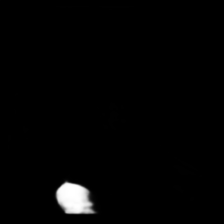}}
    \hspace{-0.18cm}
    \subfigure{\includegraphics[width=0.09\textwidth]{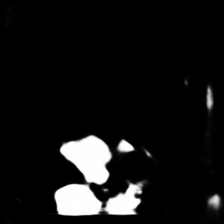}}
    \hspace{-0.18cm}
    \subfigure{\includegraphics[width=0.09\textwidth]{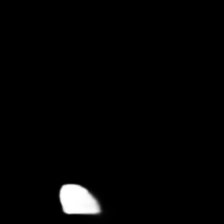}}
    \hspace{-0.18cm}
    \subfigure{\includegraphics[width=0.09\textwidth]{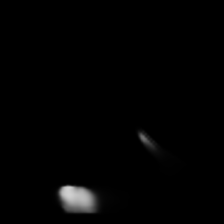}}
    \hspace{-0.18cm}
    \subfigure{\includegraphics[width=0.09\textwidth]{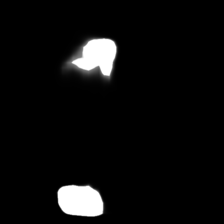}}
    \hspace{-0.18cm}
    \subfigure{\includegraphics[width=0.09\textwidth]{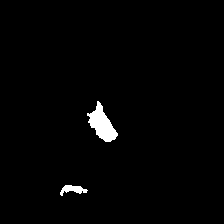}}
    \hspace{-0.18cm}
    \subfigure{\includegraphics[width=0.09\textwidth]{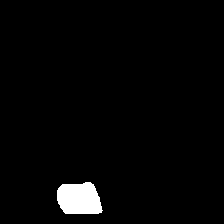}}
    \hspace{-0.18cm}
     \vspace{-0.35cm}

    \subfigure{\includegraphics[width=0.09\textwidth]{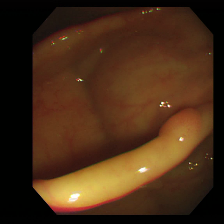}}
    \hspace{-0.18cm}
    \subfigure{\includegraphics[width=0.09\textwidth]{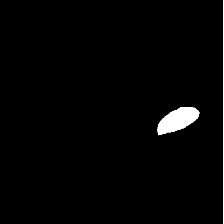}}
    \hspace{-0.18cm}
    \subfigure{\includegraphics[width=0.09\textwidth]{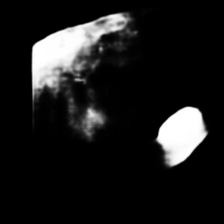}}
    \hspace{-0.18cm}
    \subfigure{\includegraphics[width=0.09\textwidth]{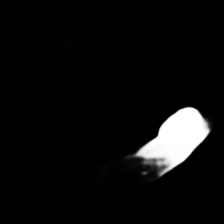}}
    \hspace{-0.18cm}
    \subfigure{\includegraphics[width=0.09\textwidth]{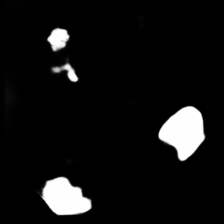}}
    \hspace{-0.18cm}
    \subfigure{\includegraphics[width=0.09\textwidth]{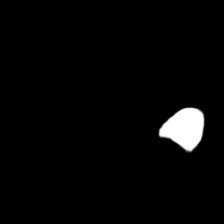}}
    \hspace{-0.18cm}
    \subfigure{\includegraphics[width=0.09\textwidth]{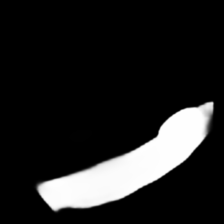}}
    \hspace{-0.18cm}
    \subfigure{\includegraphics[width=0.09\textwidth]{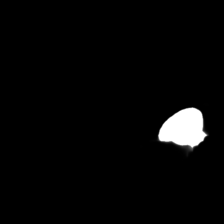}}
    \hspace{-0.18cm}
    \subfigure{\includegraphics[width=0.09\textwidth]{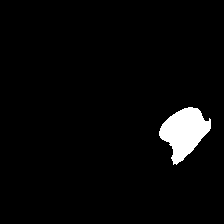}}
    \hspace{-0.18cm}
    \subfigure{\includegraphics[width=0.09\textwidth]{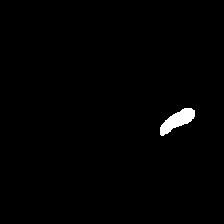}}
    \hspace{-0.18cm}
     \vspace{-0.35cm}

    \subfigure{\put(-1,-10){image}\includegraphics[width=0.09\textwidth]{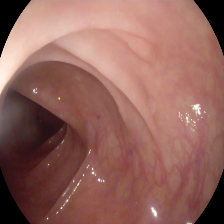}}
    \hspace{-0.18cm}
    \subfigure{\put(-1,-10){GT}\includegraphics[width=0.09\textwidth]{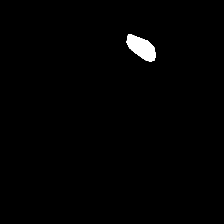}}
    \hspace{-0.18cm}
    \subfigure{\put(-1,-10){U-Net}\includegraphics[width=0.09\textwidth]{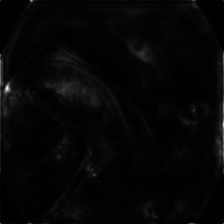}}
    \hspace{-0.18cm}
    \subfigure{\put(-1,-10){UNet++}\includegraphics[width=0.09\textwidth]{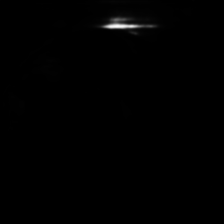}}
    \hspace{-0.18cm}
    \subfigure{\put(-1,-10){ SFA}\includegraphics[width=0.09\textwidth]{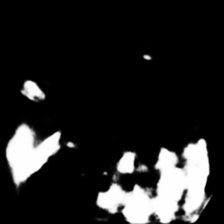}}
    \hspace{-0.18cm}
    \subfigure{\put(-1,-10){PraNet}\includegraphics[width=0.09\textwidth]{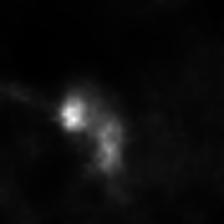}}
    \hspace{-0.18cm}
    \subfigure{\put(-1,-10){MSNet}\includegraphics[width=0.09\textwidth]{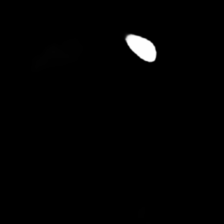}}
    \hspace{-0.18cm}
    \subfigure{\put(-1,-10){SANet}\includegraphics[width=0.09\textwidth]{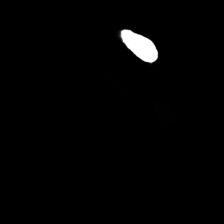}}
    \hspace{-0.18cm}
    \subfigure{\put(-1,-10){Plu nC}\includegraphics[width=0.09\textwidth]{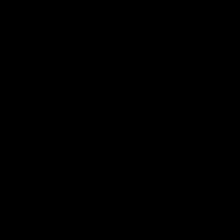}}
    \hspace{-0.18cm}
    \subfigure{\put(-1,-10){Plu wC}\includegraphics[width=0.09\textwidth]{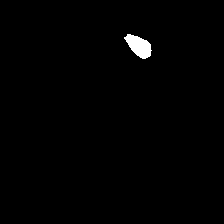}}
    \vspace{-0.35cm}

    \caption{Sample segmentation results of the benchmark models compared with PlutoNet without (Plu nC) and with consistency training (Plu wC). Images shown in the first column belong to Kvasir, ClinicDB, ColonDB, EndoScene, and Etis datasets.}
    \label{5_dataset_fig}
\end{figure}

To show the effectiveness of our decoder consistency training approach, we compared our model with and without our consistency training approach. Table \ref{ablation_data_2} shows the results of our model's performance with and without consistency training. Using our consistency training approach, we are able to reduce false positive rates and improve the segmentation results for Kvasir-SEG \cite{jha2020sessile}, ClinicDB \cite{PMID:25863519}, ETIS \cite{etis}, and EndoScene \cite{endoscene} datasets. Sample segmentation results as seen in Figure \ref{5_dataset_fig} also support these improvements.

\begin{table}
\centering
\caption{Our ablation study shows the effectiveness of our consistency training. }\label{ablation_data_2}
\begin{tabular}{|l|l|l|l|l|}
\hline
Ablation Study  & Dice & IoU & Precision & Recall \\
\hline
Kvasir No Consistency        & 0.8839          & 0.7920          & 0.9250          & \textbf{0.8315} \\ 
\hline
Kvasir With Consistency    & \textbf{0.8954}          & \textbf{0.8105}          & \textbf{0.9559}          & 0.8265  \\
\hline
ClinicDB No Consistency & 0.8964 & 0.8122 & 0.9492 & 0.8451   \\
\hline
ClinicDB With Consistency & \textbf{0.9085} & \textbf{0.8323} & \textbf{0.9515} & \textbf{0.8662} \\
\hline
ColonDB No Consistency & \textbf{0.7178} & \textbf{0.5598} & 0.7923 & \textbf{0.6556} \\
\hline
ColonDB With Consistency & 0.6935 & 0.5308 & \textbf{0.8845} & 0.5702 \\
\hline
ETIS No Consistency & 0.8042 & 0.6725 & 0.7929 & 0.8165 \\
\hline
ETIS With Consistency & \textbf{0.8296} & \textbf{0.7088} & \textbf{0.8343} & \textbf{0.8255} \\
\hline
EndoScene No Consistency & 0.8979 & 0.8147 & 0.8905 & 0.9024 \\
\hline
EndoScene With Consistency & \textbf{0.9192} & \textbf{0.8506} & \textbf{0.9221} & \textbf{0.9186} \\
\hline
\end{tabular}
\end{table}

\section{Conclusion}

With PlutoNet, we propose to learn representations that mediate between capturing abundant salient details and higher-level semantic details without being redundant. 
We enforce consistency by combining the loss of the modified partial decoder and the auxiliary decoder, which encourages the predictions of the decoders to
be consistent. PlutoNet requires far fewer parameters than its counterparts. Our experiments show that PlutoNet significantly outperforms benchmark models, particularly on unseen datasets and datasets across different domains. Although our model achieves state-of-the-art results, there are some limitations to it. By ignoring lower-level features, we are able to largely decrease redundant information, however, we might be missing tiny polyps. This is a trade-off in order to reduce the number of parameters and false positives.

%
%
%
%
\bibliographystyle{splncs04}
\bibliography{mybib}
\end{document}